\documentclass{eas}
\usepackage{graphicx}
\begin{document}

\title{The Araucaria Project. Binary Classical Cepheids in the LMC} 
\runningtitle{Eclipsing binary cepheids}
\author{Dariusz Graczyk}\address{Departamento de Astronom{\'i}a, Universidad de Concepci{\'o}n, Chile}
\author{Bogumi{\l} Pilecki}\sameaddress{1,2}
\author{Grzegorz Pietrzy{\'n}ski}\sameaddress{1,2}
\author{Wolfgang Gieren}\sameaddress{1}
\author{Piotr Konorski}\address{Astronomical Observatory of Warsaw University, Poland}
\author{Igor Soszy{\'n}ski}\sameaddress{2}
\author{Andrzej Udalski}\sameaddress{2}
\author{Alexandre Gallenne}\sameaddress{1}
%
%
\begin{abstract}
The status of our work on binary classical cepheid systems in the Large Magellanic Cloud is presented. We report on results from our follow up of two eclipsing binary cepheids OGLE-LMC-CEP-0227 and OGLE-LMC-CEP-1812. Here we presented for the first time confirmation that a third cepheid OGLE-LMC-CEP-2532 is a true eclipsing binary cepheid with a period of 800 days. Two other very good candidates for eclipsing binaries detected during OGLE-IV survey are also discussed. 
\end{abstract}
\maketitle
\section{Introduction}
The eclipsing binary systems containing classical cepheids are very rare. Although we know  tens of binary cepheids in our Galaxy (e.g. Evans \etal~\cite{eva13}), none is an eclipsing system. However extensive search for variability in microlensing surveys in Magellanic Clouds resulted in reporting of reliable eclipsing binary candidates: a double cepheid system MACHO 05:21:54.8-69:21:50 (Alcock \etal~\cite{alc95}), first overtone cepheid OGLE LMC SC16 119952 (Udalski \etal~\cite{uda99}) and two fundamental mode cepheids OGLE-LMC-CEP-0227 and OGLE-LMC-CEP-1812 (Soszy{\'n}ski \etal~\cite{sos08}). 

Deriving fundamental physical parameters of classical cepheids is an important task regarding their great significance for extragalactic distance scale. As part of the Araucaria project we made follow up of candidate eclipsing binaries in the Large Magellanic Cloud (LMC). This allowed us to confirm first known eclipsing binary cepheid (Pietrzy{\'n}ski \etal~\cite{pie10}) and to resolve so called the mass discrepancy problem in a favour of cepheid pulsation models. Our goals are summarised as follow: 1) determination of accurate dynamical masses of classical Cepheids; 2) verification of the mean cepheid radius - pulsation period relation; 3) developing of a consistent scheme for solving light and radial velocity curves of radially pulsating star in an eclipsing binary system; 4) determination of distance independent the projection factor for cepheids; 5) analysis of the limb darkening dependence on a pulsation phase.

\begin{figure}
\includegraphics[width=\linewidth]{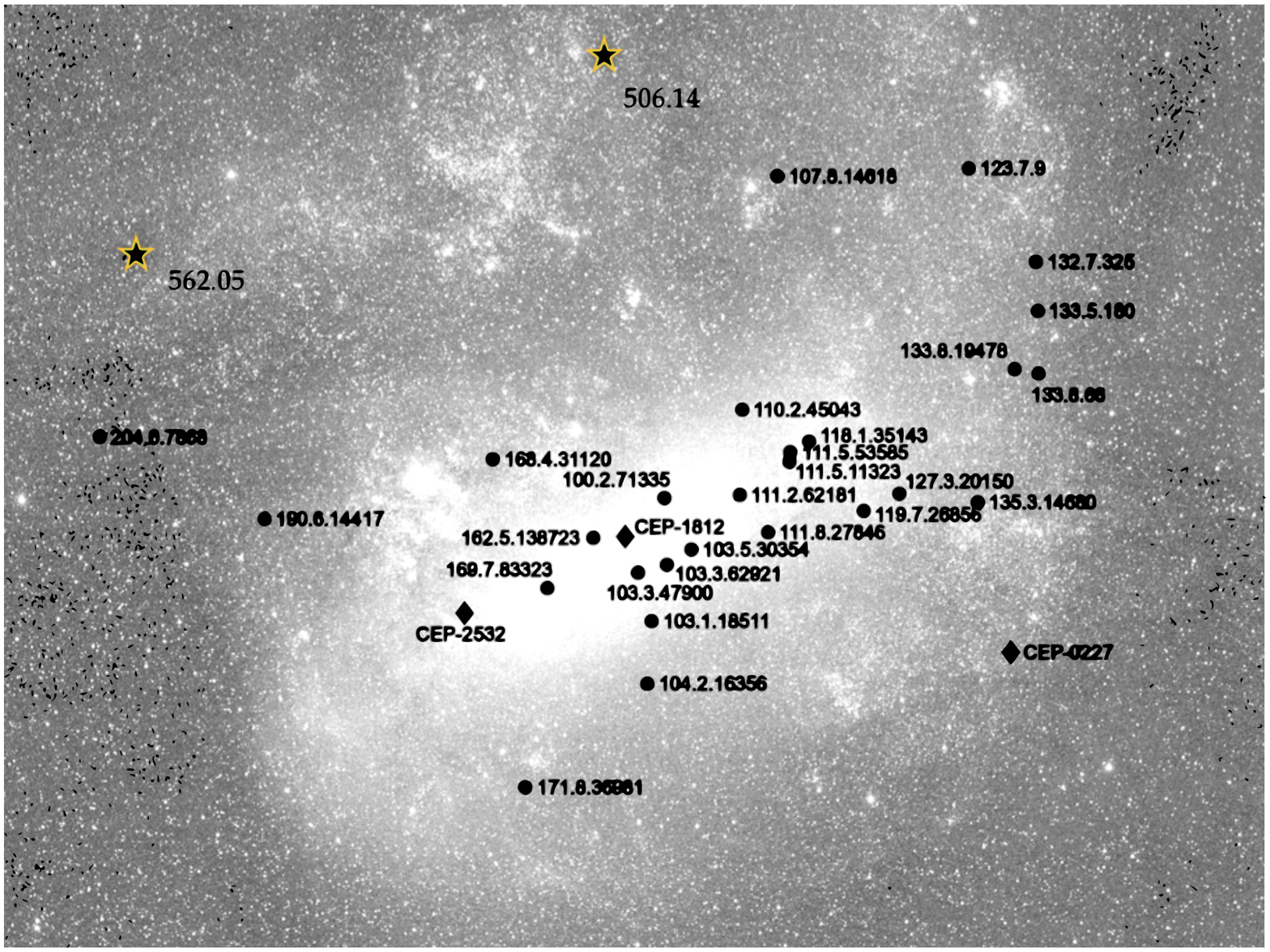}
\caption{Position of eclipsing binary classical cepheids in the LMC monitored by our project and discovered from OGLE III survey (diamonds) and OGLE-IV survey (stars). Additionally we plotted position of late type eclipsing binary stars used to determine distance to the LMC (circles). Background image is from ASAS-3 survey by Pojma{\'n}ski~(\cite{poj}). }
\label{fig1}
\end{figure}

Below we presented some results and details of our cepheids research. The sections are devoted to particular eclipsing binary systems. The Figure~\ref{fig1} presents position of our target eclipsing binary cepheids within the LMC body.

\section{OGLE-LMC-ECL-0227} 
\begin{figure}
\includegraphics[width=\linewidth]{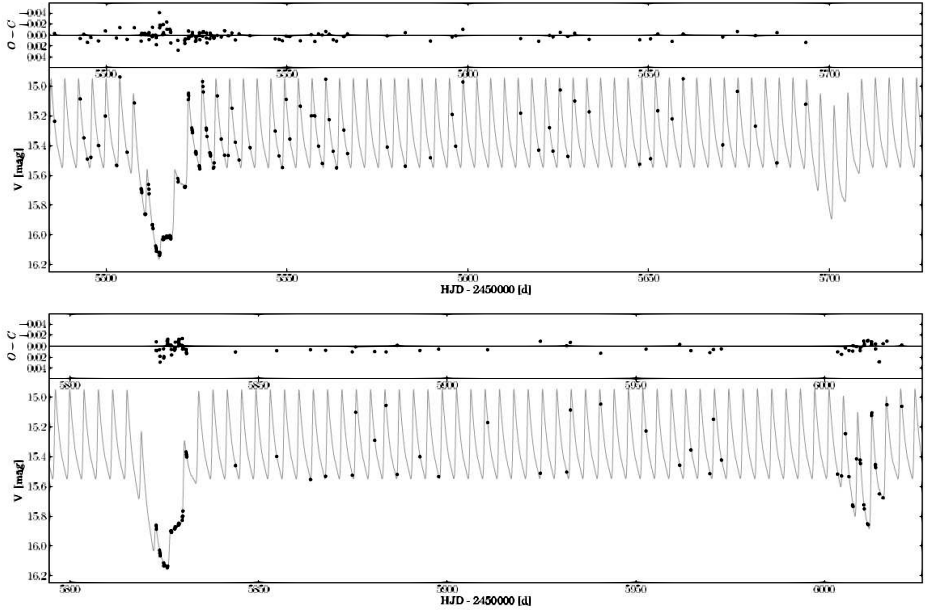}
\caption{A part of OGLE-III V-band optical light curve of eclipsing binary cepheid OGLE-LMC-ECL-0227 (points) covering two primary and secondary eclipses. The cepheid is eclipsed during deeper minimum. The continues line is our model synthetic light curve.}
\label{fig2}
\end{figure}

This relatively bright star (V = 15.2 mag) was discovered to be a classical fundamental mode cepheid by Soszy{\'n}ski \etal~(\cite{sos08}). In the same paper the cepheid was mentioned as possible eclipsing binary candidate. We confirmed it to be an eccentric double lined binary with orbital period of 309 days and deep eclipses (Pietrzy{\'n}ski \etal~\cite{pie10}). We derived physical parameters of both components and, especially, their precise masses. Subsequently we performed extensive optical and near infrared photometric and spectroscopic monitoring of this star to improve our solution. To this end we developed a novel method of light curve and radial velocity curve analysis. The new code and data allowed us to put stringent limits onto cepheid physical parameters (Pilecki \etal~\cite{pil13}). The quality of the fit is illustrated in Fig.~\ref{fig2}. 

Derived masses and radii are fully consistent with a solution reported by Pietrzy{\'n}ski \etal~(\cite{pie10}) but by a factor of two more precise. We detected anomalously large limb darkening in the pulsating component in disagreement with predictions from static atmosphere models. For more we calculated the projection factor of this cepheid which is an important parameter in Baade-Wesselink type methods of distance determination to radially pulsating stars. This direct determination is a distance and a limb darkening independent what differs our result from those obtained  for galactic cepheids from interferometry (e.g.~M{\'e}rand \etal~\cite{mer05}). The value we obtained remain in agreement with some theoretical model predictions (e.g.~Nardetto \etal~\cite{nar09}). Fundamental parameters of the cepheid component are given in Table~\ref{tab1}. The companion star has almost the same mass and luminosity like the cepheid.

\section{OGLE-LMC-ECL-1812}
\begin{figure}
\includegraphics[width=0.7\linewidth]{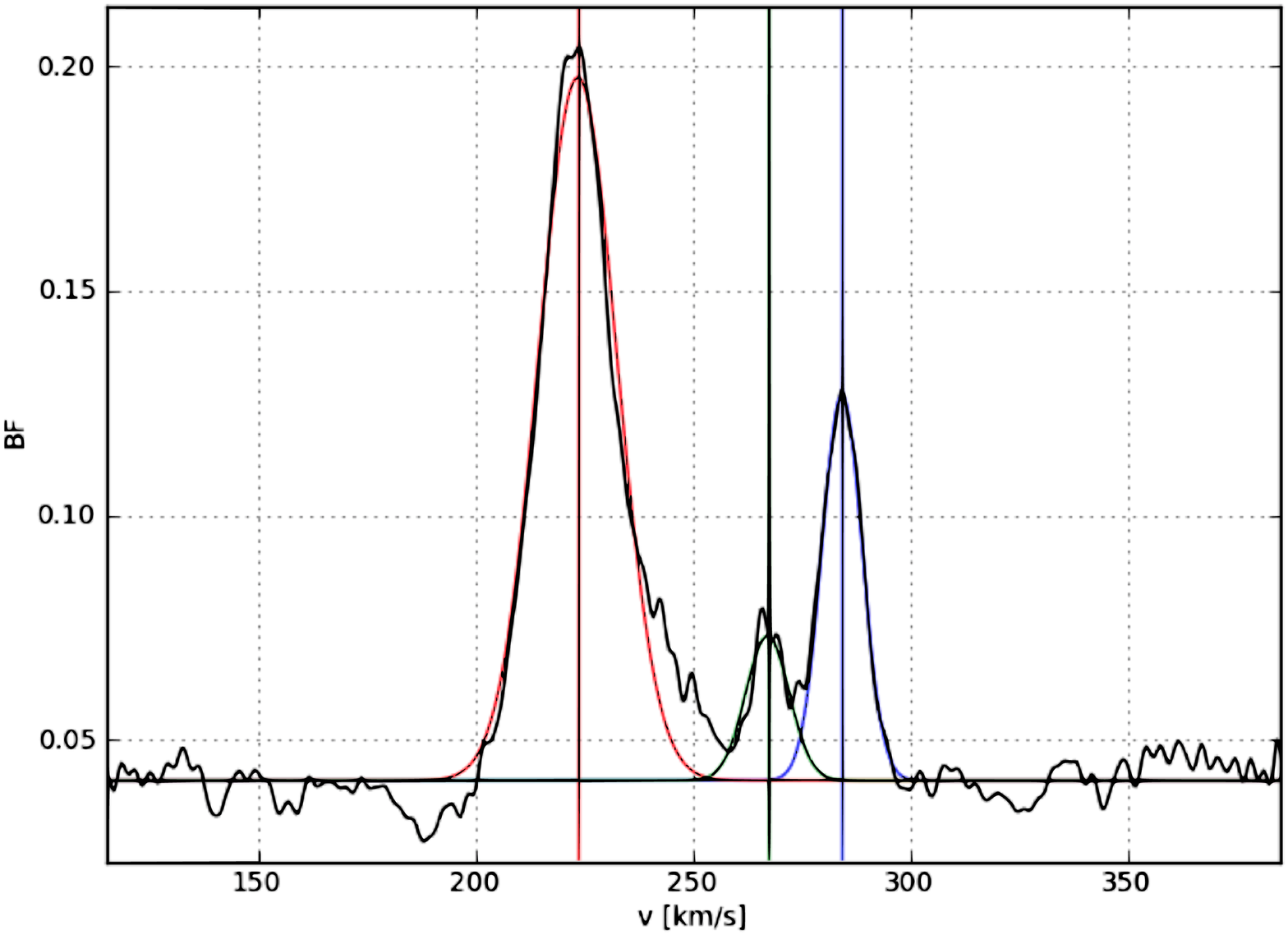}
\qquad
\includegraphics[width=0.23\linewidth]{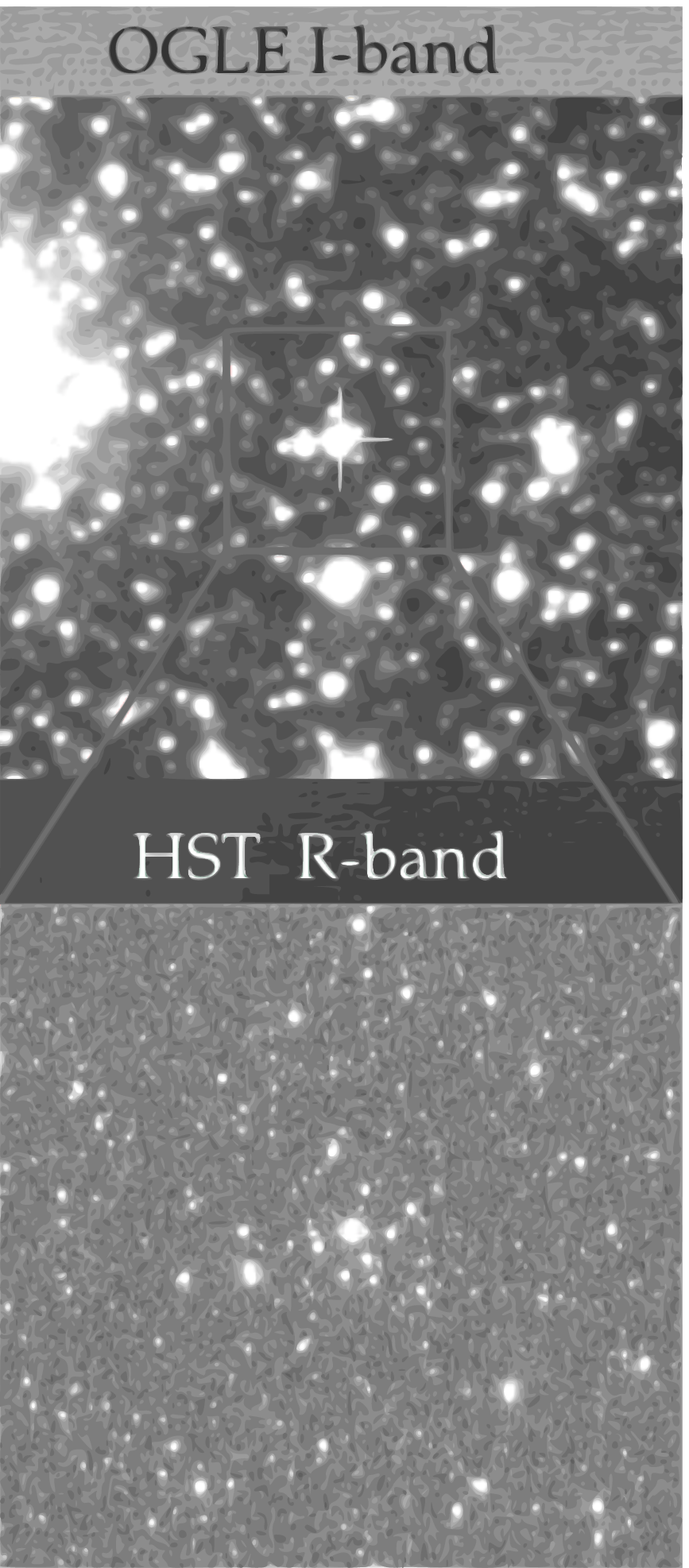}
\caption{{\it Left}: the Broadening Function of the MIKE spectrum of CEP-1812 taken close to the first quadrature. The high peak corresponds to the cepheid component and the optical blend produces a small peak at radial velocity of 265 km/s. {\it Right}: OGLE-III I-band image and close-up done with HST of a region centred on CEP-1812 (below). A clump of stars in vicinity of the cepheid system is clearly visible in R-band HST image. }
\label{fig3}
\end{figure}

The star was confirmed to be an eccentric double lined eclipsing binary with an orbital period of 552 days by Pietrzy{\'n}ski \etal~(\cite{pie11}). The pulsating component is a fundamental mode classical cepheid having shorter pulsation period than CEP-0227 - see Table~\ref{tab1}. However during analysis of this system we detected non-negligible third light. First hint of it came from spectroscopy as we observed an additional peak in the Broadening Function power distribution - Fig.~\ref{fig3}. The peak is not observed on all spectra suggesting that the third light comes from an optical blend. In our analysis we assumed that the light contribution from this blend is about 10\% of the total flux in I-band. The presence of the optical blend(s) was confirmed later through HST multicolour imaging. As the orbital inclination of the system is very close to 90 degrees the presence of the third light does not affect significantly masses reported by Pietrzy{\'n}ski \etal~(\cite{pie11}). Thus the cepheid component is only the second one with mass known with accuracy better than 3\%. However radii and luminosities of both component are a subject of some uncertainty up to 5-10\%. This system is somewhat puzzling because the cepheid has much larger mass (and luminosity) than the companion star, however we meet both these stars in a short evolutionary phase of giant evolution. Currently we work on improving our previous solution and deriving the p-factor for this cepheid.  

\section{OGLE-LMC-CEP-2532}
\begin{figure}
\includegraphics[width=\linewidth]{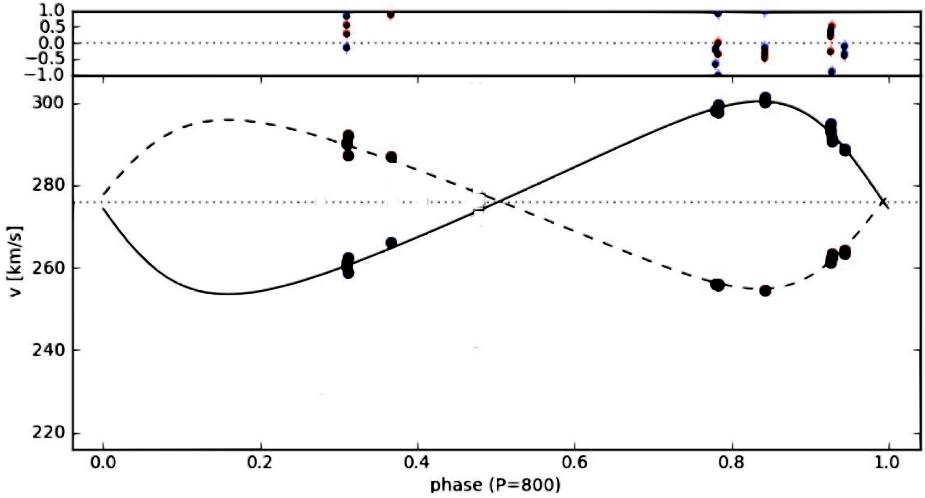}
\caption{The radial velocities and a model orbital motion of components of the system CEP-2532 after subtracting cepheid's pulsation velocities - points along continuous line. The points along dashed line signify radial velocities of red giant companion star.}
\label{fig4}
\end{figure}

This star contains a first overtone cepheid in a eccentric double lined eclipsing system having an orbital period of 800 days. We confirm here that both stars are gravitationally bounded by detecting their mutual orbital motion - see Fig~\ref{fig4}. Because of the eccentricity and a position of the orbit we observe only one minimum in this system when a red giant companion is eclipsing the cepheid star. Our photometric campaign to detect secondary minimum around expected orbital phase 0.55 gave null result. It is interesting that the companion star is in fact more massive and more luminous than the cepheid.  

\section{Cepheids LMC562.05.9009 and LMC506.14.8910} 
\begin{figure}
\includegraphics[width=0.4\linewidth]{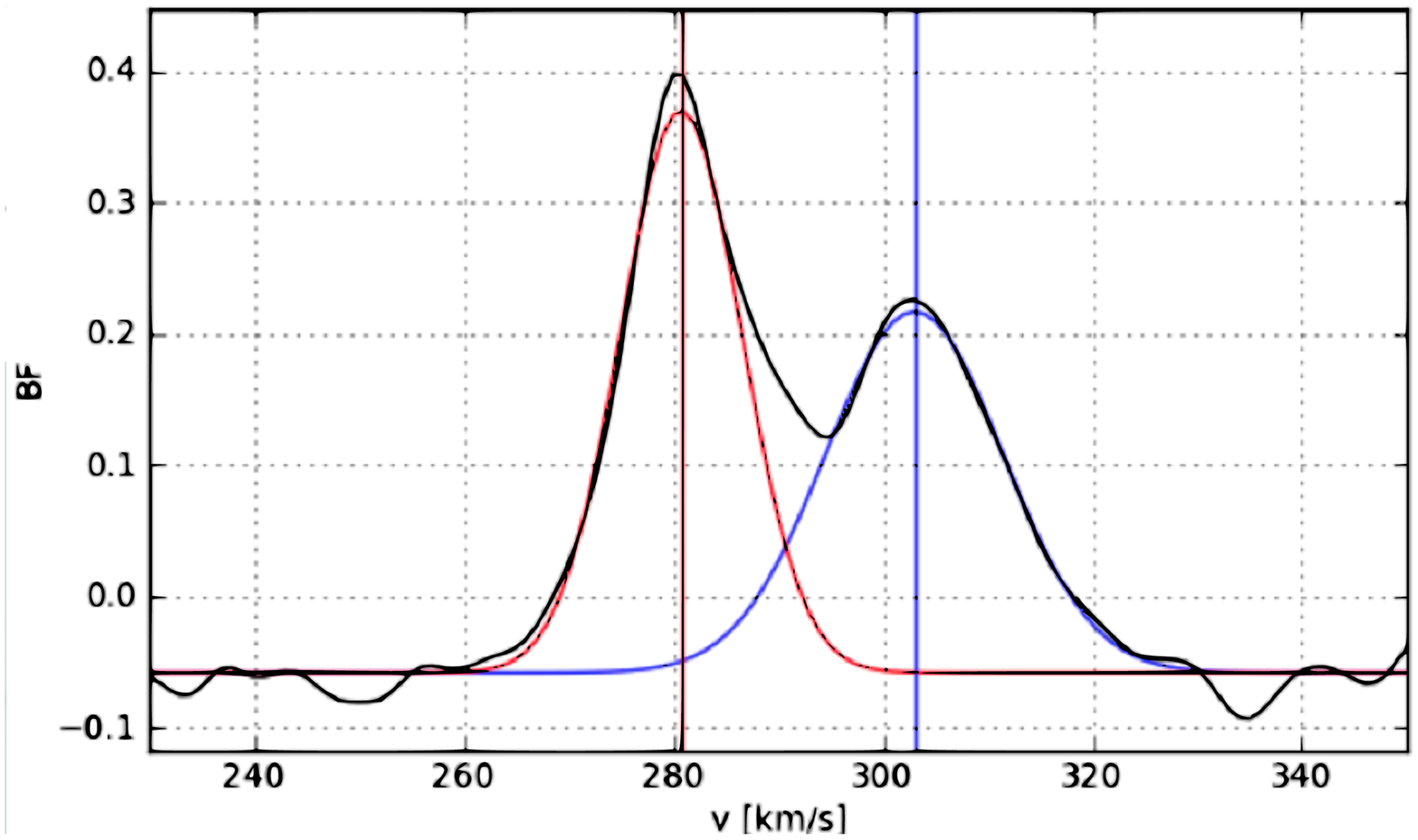}
\qquad
\includegraphics[width=0.55\linewidth]{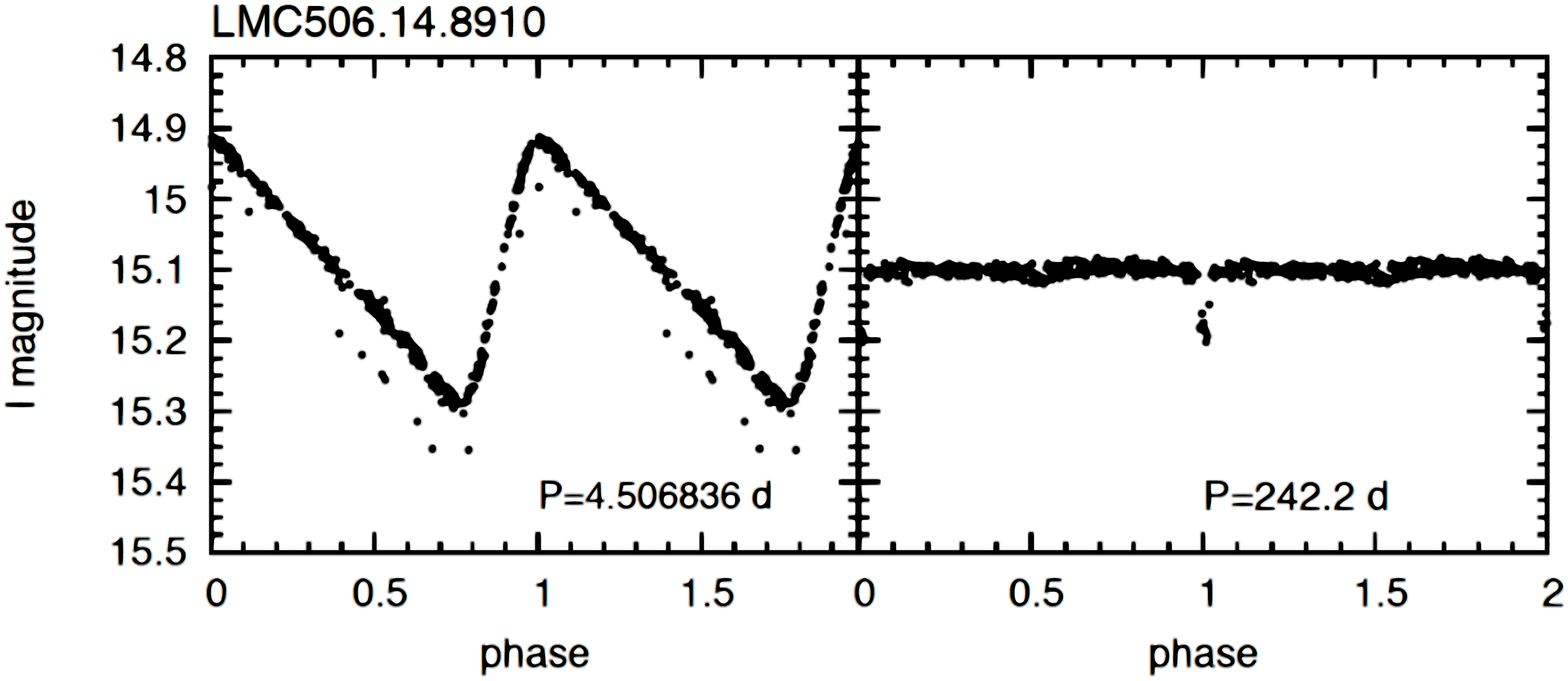}
\caption{{\it Left}: the Broadening Function of the HARPS spectrum of LMC562.05. The two high peaks correspond to the cepheid component and its probable physical companion.  {\it Right}: OGLE-IV I-band light curve of LMC506.14 phased with pulsation period (the left subplot) and with supposed orbital period. }
\label{fig5}
\end{figure}

Star LMC562.05.9009 is reported by Soszy{\'n}ski \etal~\cite{sos12} as candidate eclipsing binary cepheid based on results from the Gaia South Ecliptic Pole search by OGLE-IV survey. The cepheid star pulsates in fundamental mode. We monitor this star extensively with spectroscopy and photometry. Up to now we observed only two deep eclipses of comparable depth (primary and secondary minimum) and thus we cannot calculate orbital period. We collected some spectra which show two strong peaks in the Broadening Function power (Fig.~\ref{fig5}) suggesting that this is a true binary. 

Star LMC506.14.8910 was found to be a candidate eclipsing binary cepheid during search for variable stars in OGLE-IV survey. The cepheid also pulsates in fundamental mode, however eclipses are shallow. Especially we see only a trace of a possible secondary eclipse at orbital phase 0.53 - see Fig.~\ref{fig5}. At the moment we have no information about  a presence of a companion star in spectra.  

\begin{table}[t]
\caption{Physical parameters of the LMC classical Cepheids in eclipsing binaries}
\label{tab1}
\begin{tabular}{|c|r|r|r|r|r|}
\hline
Parameter & CEP-0227 & CEP-1812 & CEP-2532 & LMC562 & LMC506 \\ \hline
Orb. period (d)  & 309 & 552& 800& $>820$ & 242 (?)\\
Pul. period (d) & 3.98&  1.31& 2.04 & 2.99 & 4.51 \\
Pul. mode  & F& F& 1O& F & F \\
Mass  (M$_{\odot}$) & $4.17\pm0.03$ & $3.74\pm0.06$&$\sim2.4^a$& ?$^b$ & ?\\
Mass ratio &$0.99\pm0.01$& $0.71\pm0.02$ &$1.15\pm0.10$& ?& ?\\
Radius (R$_{\odot}$) &$34.9\pm0.3$ & $17.4\pm0.9$ & ?& ? & ?\\
p-factor & $1.21\pm0.03$ & $\sim1.3$ & ? &?  & ?\\
Status & confirmed & confirmed & confirmed & probable & probable \\ \hline
\end{tabular}
{{\it Notes:}\\
  $^a$ Assuming the orbital inclination of 90 deg.\\
  $^b$ ? - not directly determined yet.}
\end{table}

\section{Summary}
During a few last years we were able to significantly improve the knowledge about physical parameters of classical cepheids by observing and analysing two eclipsing binary cepheids in the LMC. At least one more eclipsing binary system OGLE-2532 is confirmed. Our future plans consist of 1) performing similar analysis for CEP-1812 like we did in the case of CEP-0227, 2) follow up of two new candidate eclipsing binaries, 3) confirmation of binary nature of double eclipsing cepheid OGLE-LMC-CEP-1718 and 4) improving our understanding of the Baade-Wesselink method applied for cepheids. 
\section*{Acknowledgements:} 
We gratefully acknowledge financial support for this work from the Polish 
National Science Center grant MAESTRO DEC-2012/06/A/ST9/00269.

\end{document}